\documentclass[10pt,twocolumn]{article}
\usepackage[letterpaper,margin=0.70in,columnsep=0.25in]{geometry}
\usepackage[T1]{fontenc}
\usepackage{lmodern,microtype,amsmath,amssymb,booktabs,graphicx,enumitem,needspace}
\usepackage[numbers,sort&compress]{natbib}
\usepackage[colorlinks=true,allcolors=blue]{hyperref}
\usepackage{url,seqsplit}
\setlist{nosep,leftmargin=*}
\title{DFIP: Decision-Focused Inference Path\protect\\for System One Models}
\author{Ismail Pelaseyed \qquad Alan Zabihi\\Superagent Technologies Inc.}
\date{October 19, 2026}
\begin{document}
\maketitle
\begin{abstract}
System One models (also known as decision models) return small structured answers, yet their request paths can retain host dispatch, GPU launch overhead, and full-vocabulary projection. Here, we present DFIP (Decision-Focused Inference Path), an empirical systems study applying established inference techniques to four pinned models: PPLX-Decider, Liquid d1-3B, Clef-Flash, and GLiNER2.5-Decide. Its contribution is a paired native-baseline evaluation, exposed-answer audits, and separate arrival measurements. The implementation applies selective CUDA graph replay, candidate-token projection where the released readout permits it, and schema-compatible length-aware batching at compatible model boundaries. With optional native Qwen kernels installed, DFIP achieves short Boolean resident speedups over strengthened native inference of 2.47$\times$ for PPLX-Decider over five containers and 1.45$\times$ for Clef over four of five planned containers; Clef's fifth container fails in native preflight. Liquid is effectively tied with its released compiler setting on the short Boolean workload. Replay supplies most of Liquid's initial eager-baseline gain; projection alone gives little full-request improvement. In separate one-worker GLiNER pilots, length-aware batching is 1.57$\times$ faster than eager native FIFO batching and 1.40$\times$ faster than compiled native FIFO batching on a diverse resident trace, while passing the exposed-label and confidence audit. The compiled arrival replay instead has worse shared-scheduler p95 than native batching. Audits of 3,024 decisions per model preserve all 1,112 returned choices for Decider and Clef, but Clef changes scores, probabilities, and sampled serving confidence thresholds. Diverse external arrivals reverse fixed-shape replay gains and expose overload losses and a replay-path CUDA error. The results identify workload-specific acceleration and its limits, rather than a universal optimizer or production readiness.
\end{abstract}

\Needspace{8\baselineskip}
\section{Introduction}
DFIP (Decision-Focused Inference Path) is an implementation and empirical study of established execution techniques in four pinned System One models: PPLX-Decider, Liquid d1-3B, Clef-Flash, and GLiNER2.5-Decide. Our contribution is a four-model characterization of when those techniques improve on strengthened or released native execution, what observable answers they preserve, and how their gains change under arrivals. Table~\ref{tab:overview} gives the cross-model overview; the experimental phases remain separate.

\begin{table*}[t]
\centering\small
\setlength{\tabcolsep}{3pt}
\begin{tabular}{p{0.15\textwidth}p{0.23\textwidth}p{0.19\textwidth}p{0.34\textwidth}}
\toprule
Checkpoint & Native comparator & Native/DFIP & Evidence and limit \\
\midrule
PPLX-Decider & Native with CUDA extensions & 2.47 [1.88, 3.26] & Short Boolean; five completed containers \\
Clef-Flash & Native with CUDA extensions & 1.45 [1.33, 1.71] & Short Boolean; four of five containers; native preflight failure and confidence drift retained \\
Liquid d1-3B & Released compiled native & 1.01 [0.97, 1.09] & Short Boolean; three containers; effectively tied \\
GLiNER2.5-Decide & Compiled native FIFO batching & 1.40 & Diverse resident trace; one worker; worse arrival p95 \\
\bottomrule
\end{tabular}

\caption{Four-model overview of strengthened and released native comparisons. Native/DFIP divides resident latency or trace completion time by its DFIP counterpart; above one favors DFIP. Decider, Clef, and Liquid use paired container medians and 95\% bootstrap intervals. GLiNER uses a diverse resident trace in one worker without an independent-worker interval. The rows use different workloads and are not a ranking or pooled estimate. Serving and answer-fidelity limits are evaluated separately.}
\label{tab:overview}
\end{table*}

System One models, also known as decision models, serve routing, policy classification, and structured scoring tasks. An application supplies a state and a small set of questions, and the model returns labels, confidences, or scores. The output is short, but the computation can still include repeated Python dispatch, many small GPU launches, and a projection over tokens unused by the released readout. Reducing this work can lower resident latency and the GPU time required for a workload. The relevant result is an improvement over an efficient native implementation with the application's observable answer contract preserved.

CUDA graphs, compilation, constrained selection, and batching already provide mechanisms for this purpose~\citep{cudagraphs,pygraph,sglang,vllm}. Applying them to System One checkpoints raises practical questions. A graph captured for one shape may be expensive to replace under changing traffic. A smaller projection can preserve a mathematical identity but change floating-point reductions. Batching can reduce total work while increasing a request's wait. A winning label may stay fixed while its confidence crosses an application threshold.

The models expose different released engines and answer heads. We preserve those interfaces and apply established execution techniques at compatible boundaries. Our objective is acceleration; output checks determine whether a measured improvement is usable under the evaluated contract. Native fallback by itself is not a speed contribution.

We ask three questions. \textbf{RQ1:} Which request paths improve over stronger native baselines, and which components produce the gain? \textbf{RQ2:} Do the accelerated paths preserve exposed answers, including confidence and scores? \textbf{RQ3:} Do resident gains persist under changing arrivals? The study yields three lessons:
\begin{itemize}
\item \textbf{Baseline and workload determine the gain.} Replay improves repeated short requests for Decider and Clef; released compilation closes Liquid's eager-baseline gap. GLiNER retains a resident batching gain in one exploratory worker.
\item \textbf{Answer fields constrain usable acceleration.} Preserving categorical choices can still change confidence-based routing, as Clef's literal-field and serving audits demonstrate.
\item \textbf{Resident speed does not establish serving speed.} Capture costs and queues reverse the original replay gains on diverse external traffic; GLiNER's compiled resident gain does not improve its tested arrival tail.
\end{itemize}
The implementation and retained positive, negative, and failed outcomes support these finite-sample conclusions. We do not propose a new CUDA graph algorithm or claim transfer to all System One models.

The public repository is \url{https://github.com/superagent-ai/dfip}. Python implements adapters, scheduling, validation, and remote orchestration; Triton~\citep{triton} implements an experimental candidate projection. Checkpoints and corpora stay on Modal. The lightweight local package contains neither weights nor accelerator dependencies. Appendix~\ref{app:artifact} identifies the frozen analysis revision, execution-source index, and dependency profiles.

\section{Model Contracts and Scope}
An input contains a state $s$ and typed questions $Q$. For the three generative-backbone checkpoints, a categorical question returns a finite distribution and selected answer, a Boolean question returns confidence in the affirmative answer, and an ordinal question can expose a distribution and expected score. Released prompt format, criterion ordering, token aliases, temperature, and answer conversion are part of the contract.

Liquid's d1-3B uses an LFM-based engine with a packed trunk/branch path for multiple questions. Clef-Flash uses a Qwen backbone and joint schema head. PPLX-Decider uses its released Qwen-based decision engine~\citep{liquid,clef,decider}. GLiNER2.5-Decide uses released schema-conditioned classification~\citep{glinerdecide,gliner2}. Its tested API exposes a winning label and confidence. Six score labels in GLiNER fixtures are categorical buckets; we do not manufacture an ordinal expectation or an unexposed probability vector. Its tested Boolean schema is likewise a two-label classification task.

The study concerns text inputs, unchanged weights, and released/default precision. It does not compare model task quality, train, quantize, or distill a checkpoint. The four-model scope is a focused case study selected after exploratory development, rather than a random sample or unseen-architecture evaluation. Trials for these checkpoints remain in the artifact, including negative results and exclusions. Separate model experiments are outside this paper's claims.

\subsection{Candidate projection and numerical fidelity}
Let $h\in\mathbb{R}^{d}$ be the hidden vector at an answer position and $z=Wh$ the vocabulary logits. For a released readout that groups answer-token aliases $A_c$, consider
\begin{equation}
 r_c=\max_{t\in A_c}z_t,\qquad
 p_c=\frac{\exp(r_c/\tau)}{\sum_j\exp(r_j/\tau)}.
\end{equation}
If $U=\bigcup_c A_c$, computing $W_Uh$ preserves the needed logits in real arithmetic. A vocabulary-wide normalization constant cancels when selected scores are normalized over candidates. This reduction is applicable only when the readout consumes the covered token groups. Arbitrary generation, full-vocabulary confidence, vocabulary-wide top-$k$, or an incomplete alias set require different analysis. A dedicated classification head already avoids this projection.

The identity does not imply finite-precision equivalence. If $\|p-\hat p\|_\infty\leq\epsilon$, an argmax with margin greater than $2\epsilon$ is preserved. A threshold at $a$ is preserved only if $|p_c-a|>\epsilon$. For ordinal levels $v_c$, expected-score drift is at most $\sum_c |v_c|\,|p_c-\hat p_c|$. Separately rounded or computed API fields require direct comparison. We report labels, probabilities, literal confidence, and literal scores when the native API exposes them. Label agreement alone is weaker than preserving a confidence-based application policy.

\section{Execution Paths}
\subsection{Selective replay}
An adapter invokes the released engine at an immutable checkpoint revision. DFIP wraps a compatible tensor boundary with a CUDA graph, static input buffers, and a shape key containing dimensions, strides, dtype, device, and supported scalar metadata. Replays copy the current request into the buffers; the cache stores computation rather than predictions. Calls using mutable buffers are serialized. Clef and Decider use replay around the native backbone without substituting a candidate head. Liquid combines replay with its candidate projection.

The original tested policy admits a shape after three encounters and uses a four-entry least-recently-used cache. All three checkpoints limit capture to 512 tokens; Clef and Decider additionally limit it to one forward row. Decider includes each sequence's non-padding token count in the graph-cache key because its hybrid recurrence launch metadata depends on those lengths. Tensor shape alone cannot establish equivalent control flow. Unsupported paths invoke native inference. These restrictions are part of the measured system, so one admitted short request cannot establish a model-wide speedup.

Allocation lifetime and capture work matter. Earlier Liquid and Clef experiments abort after illegal-memory-access errors. A follow-up synchronizes before clearing or evicting graphs; a synthetic 60-call changed-shape test passes, but this does not prove model-level reliability. The strengthened Decider serving study later encounters a replay-path CUDA error. Asynchronous reporting prevents assigning a kernel cause from those tracebacks alone. Reviewed public inference defaults retain native execution; research entry points expose replay.

\subsection{Candidate readout for Liquid}
The Liquid adapter derives answer-token groups from the released tokenizer and readout. A projection computes the verified union using either a Triton reduction or a PyTorch linear operation backed by cuBLAS. Extending the union invalidates existing captures. To preserve the engine's tensor interface, the prototype reconstructs a vocabulary-shaped tensor with uncomputed positions set to negative infinity. It removes projection arithmetic but retains allocation, scatter, and readout overhead.

Exploratory RMSNorm replacements changed winning labels for Liquid, Clef, and Decider and are not enabled. A local reduction test cannot establish full-model answer preservation: numerical changes can propagate through the backbone and calibrated readout. Released compiler options are tested as baselines rather than presumed slower than custom code.

\subsection{Bounded capture admission}
After varied-serving results, a separate post hoc prototype replaces eviction with a bounded capture budget. It records the last 128 eligible calls and waits for at least 32 observations. A shape requires eight matching hits and frequency at least 12.5\%; each configuration permits at most four capture attempts, including failures. Admitted graphs are pinned. Other shapes execute natively; a new worker or reconfiguration resets the budget.

This policy bounds admission work but does not replace stale shapes when popularity changes. Its tested API accepts one categorical question on one record, with inherited single-row and 512-token limits. It still serializes GPU access and cannot recover a CUDA context damaged by illegal memory access. The one-worker pilot is separate from the original policy.

\subsection{Schema-compatible length-aware batching}
The GLiNER experiment uses its released single and batch calls. An adapter supplies an ordered schema key and input-length estimate. DFIP includes the oldest pending request, prefers compatible requests of similar length, caps batches at eight, caps estimated padded tokens at 8,192, and bounds the longest/shortest length ratio to four. It restores request order and changes neither precision nor the classifier head.

A finite calibration trace admits sizes two, four, and eight separately per ordered schema. Every compared literal field and structure must match. Numerical tolerance is 0.001, with no exposed confidence or probability crossings at 0.1, 0.25, 0.5, 0.75, or 0.9. Each size must exceed 1.10$\times$ serial native speed across three diagnostic timings. Partial batches use only an independently admitted size; new schemas remain singletons. Passing this finite gate does not guarantee arbitrary future inputs and is not a per-request shadow comparison.

The stronger comparator already groups identical ordered schemas into native FIFO batches of up to eight. The incremental question is therefore length-aware selection over native batching. A two-millisecond accumulation window is used in the short arrival replay. These are established scheduling mechanisms integrated with a released decision API, rather than a new general batching algorithm.

\section{Experimental Method}
\subsection{Hardware and native baselines}
Liquid, Clef, and GLiNER run on NVIDIA L40S; PPLX-Decider runs on NVIDIA B200. Each experiment uses one GPU per worker. Strengthened and follow-up jobs use one study GPU at a time. The artifact records GPU identity, driver, CUDA, package versions, container identity, load time, source commit, and mode order. Fresh containers need not use independent physical hosts.

The main replay image uses PyTorch 2.14.0, CUDA 13.0, Triton 3.8.0, and Transformers 5.17.0. For Clef and Decider, it installs \texttt{causal-conv1d} 1.7.0 and \texttt{fla-core} 0.5.2 for released Qwen linear attention~\citep{qwenkernels}. SDPA and Decider's noncausal attention semantics remain intact. Original fallback-image samples are not pooled with this stronger baseline. Changes between images and containers cannot isolate the effect of installing an extension.

Liquid is compared with its released runner's \texttt{torch.compile(mode="reduce-overhead")} setting~\citep{torchcompile}. GLiNER uses \texttt{gliner2} 2.1.0 and released/default FP32 tensors, first eagerly and then with the library's \texttt{compile()} entry point. The compiled GLiNER follow-up is post hoc and reuses the pilot fixtures. These are concrete native comparators, not an exhaustive tuned-serving ceiling.

\subsection{Resident measurements and statistical unit}
For Clef and Decider the strengthened protocol plans five single-use container rounds. Nine workloads cover short Boolean and categorical decisions, three questions on one state, medium and long repeated-log contexts, candidate widths two/eight/sixteen, and eight requests in one batch. Each mode receives 50 timed calls after four admission/warmup calls. Mode and workload orders are randomized within rounds. GPU synchronization brackets resident API timing, including encoding, execution, and native readout; transport and startup are excluded.

The statistical unit is the container round. Each paired speedup is native median latency divided by DFIP median latency; values above one favor DFIP. We pair within-container medians and report the geometric mean of these ratios. The 95\% percentile bootstrap interval resamples paired log ratios 10,000 times; 50 calls are not independent deployment replications. Failed rounds are retained without imputation. Small samples and conditional estimates after attrition limit interpretation.

Liquid's compiler comparison uses three containers and three short workloads. An earlier three-retained-round component ablation compares eager native, replay alone, projection alone, and their combination. The dynamically extended answer union has 40--42 tokens during measurement. This component evidence is not pooled into the compiler comparison. Earlier Liquid and Clef studies each stopped after three rounds following GPU errors; they cannot supply missing repetitions of later protocols.

GLiNER begins with 29 native conformance records, repeated native calls, eight resident cells with 20 warm samples, and single batch probes. The scheduler calibrates on 96 records and evaluates separate 96-record recurring-schema and diverse traces. The latter changes types, option counts, state lengths, and some criterion ordering. Five paired randomized warm timings per trace run in one worker, providing within-worker medians rather than independent-worker intervals. Calibration and evaluation use different state templates, but native architecture and profiling evidence is available before policy design.

\subsection{Answer audits}
For Decider and Clef, 400 cases from pinned \texttt{LocalLLaMA/typed-decisions} supply 2,000 questions. This test set was used during development. Another set derives 512 records and 1,024 decisions from 256 BANKING77 test utterances~\citep{banking}. Widths cycle through two, four, eight, and sixteen and include the gold intent. Each utterance has a single-choice and a choice/Boolean/ordinal view. Boolean targets alternate between matching and nonmatching intents. Ordinal questions are diagnostic without validated score annotations. The correlated views are not independent quality examples, and training overlap is unknown.

A separate instrumented audit compares literal returned choice, score, and confidence plus exposed probabilities on all 3,024 decisions. Clef rounds slots after selecting the categorical answer and computes its score from unrounded probabilities; reconstructing these fields cannot establish literal API equality. The audit uses one container per model with fixed native-then-DFIP order and is separate from timing replication.

Liquid checks 128 changing, unlabeled categorical requests per compiler round, varying content, length, and width; these checks do not cover every Boolean/ordinal timing schema. GLiNER compares labels and confidence with serial native on resident traces and completed arrival requests. Its original pilot archives calibration comparison summaries; individual calibration answers and full serving references are additionally retained in the compiler follow-up. Fidelity results remain finite-sample statements.

\subsection{External arrivals and short resident replay}
The strengthened serving protocol plans three application/container rounds for Decider and Clef. An external workstation with Swedish network egress sends authenticated Modal RPC requests at fixed open-loop rates of five, fifteen, and thirty per second for 60 seconds each. Content, twelve recurring state-repetition counts, and widths two/eight/sixteen change deterministically. This trace can exceed the four-entry cache. Its payloads differ from the resident short Boolean request and do not measure one fixed captured shape.

Workers accept up to 64 concurrent RPC inputs but serialize GPU execution to protect graph buffers. The client allows 128 requests in flight and drops excess arrivals. Each phase drains before the next. Latency starts at scheduled arrival, including dispatch delay, transport, queueing, and response. Initialization and 48 admission calls are separate. Persistent artifacts mean initialization is not an empty-cache download benchmark. Drops and errors remain in the offered denominator; quantiles concern successful requests. Matched successful outputs are audited, with unmatched requests counted separately.

The bounded-admission pilot uses one validated worker each for Decider and Clef in the stronger image. Each mode receives the diverse trace, then a synthetic mix of 60\% recurring short requests and 40\% diversity at the same rates and durations. Content still changes; admission state persists across profiles. Decider runs native first, corrected Clef adaptive first. Clef's first attempt changes worker identity during traffic and finishes on a replacement in native mode. The whole comparison is retained and excluded. An amendment adds worker/mode checks and increases the idle window from ten to 300 seconds before a corrected trial; the replacement cause is unknown.

GLiNER instead receives short resident replays of 256 requests per trace. The recurring trace offers 25/s; the diverse trace alternates 32-request groups at 80 and 20/s. A 128-request queue, two-second deadline, and two-millisecond batch window are fixed. Scheduled-arrival latency includes DFIP length estimation. Reference evaluation warms fixtures beforehand. Arrival windows last approximately eight to eleven seconds without HTTP transport. These results are distinct from the external 60-second phases and do not constitute a long-duration service test.

\section{Acceleration Results}
\subsection{Replay against stronger native execution}
Table~\ref{tab:optimized-native} compares DFIP with strengthened native inference on short Boolean requests, with optional native Qwen kernels installed in both paths. DFIP gives a 2.47$\times$ resident speedup over this native baseline for Decider across five completed container rounds, from a geometric-mean native median of 74.54\,ms to 30.14\,ms. For Clef, DFIP gives a 1.45$\times$ resident speedup over strengthened native inference across four retained container rounds, from 51.69\,ms to 35.68\,ms; its fifth container fails during native reference collection before timing or custom capture. Errors are therefore not confined to replay, but the asynchronous trace does not identify a kernel cause. Clef's interval is conditional on retained rounds.

\begin{table*}[t]
\centering\small
\begin{tabular}{lrrrr}
\toprule
Checkpoint & Rounds & Native (ms) & DFIP (ms) & Ratio [95\% CI] \\
\midrule
PPLX-Decider & 5/5 & 74.54 & 30.14 & 2.47 [1.88, 3.26] \\
Clef-Flash & 4/5 & 51.69 & 35.68 & 1.45 [1.33, 1.71] \\
\bottomrule
\end{tabular}

\caption{Short Boolean resident comparison with native CUDA convolution and FLA bindings installed. Speedup is strengthened-native median latency divided by DFIP median latency. Latencies are geometric means of within-container medians. Counts are completed/planned rounds; intervals bootstrap paired container medians.}
\label{tab:optimized-native}
\end{table*}

\begin{figure*}[t]
\centering
\includegraphics[width=0.98\textwidth]{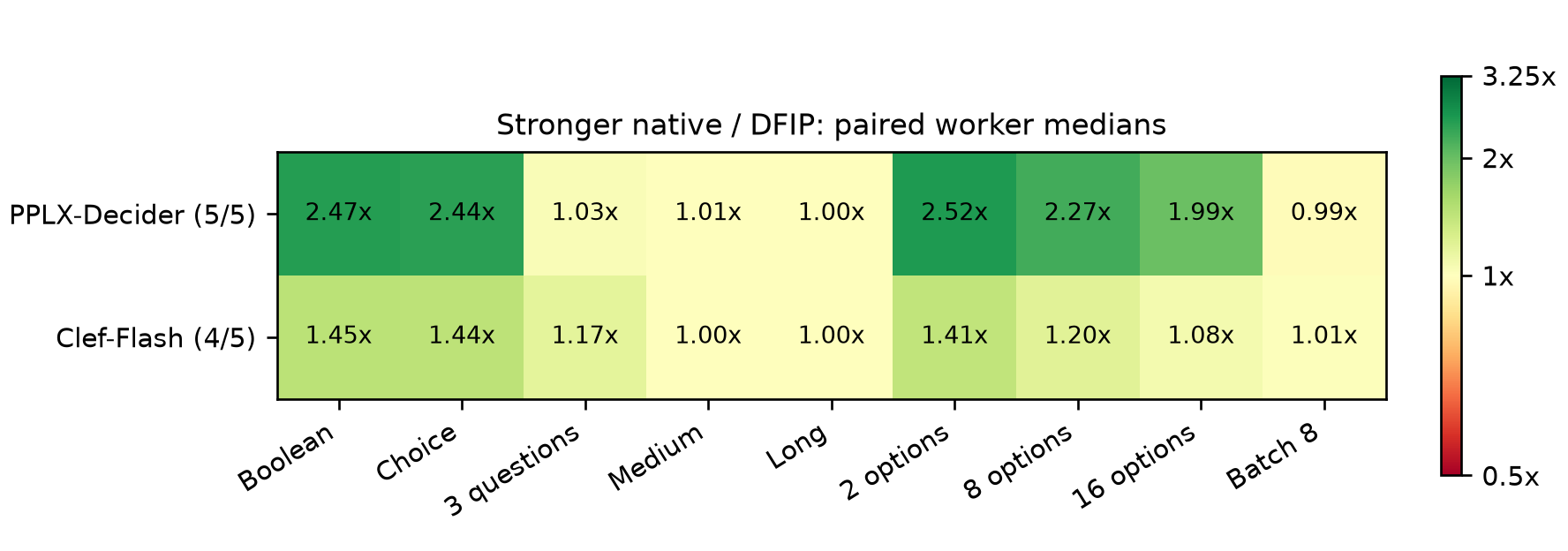}
\caption{Paired speedup across nine strengthened resident workloads, defined as strengthened-native median latency divided by DFIP median latency. Clef uses four completed rounds and Decider five. Values below one indicate slower DFIP execution; unsupported paths use native inference. Conditional intervals and per-call samples accompany each cell.}
\label{fig:optimized-workloads}
\end{figure*}

Figure~\ref{fig:optimized-workloads} shows workload dependence. Gains concentrate in repeated short single-decision requests. Long contexts and request batches use native fallback; Decider's multi-question path also falls back. Its native short Boolean medians range from 47.22 to 114.56\,ms across containers. Pairing avoids dividing unrelated runs but cannot remove mode-order, clock, or allocator-history effects. No model-wide constant speedup is justified.

\subsection{Liquid: native compilation closes the eager gap}
All three Liquid compiler containers complete. Table~\ref{tab:compiler} gives 1.01$\times$ for short Boolean requests, with an interval spanning one. DFIP is slightly slower for categorical requests and slightly faster for three questions. No general latency advantage over released compilation is established.

\begin{table*}[t]
\centering\small
\begin{tabular}{lrrrr}
\toprule
Workload & Rounds & Compiled (ms) & DFIP (ms) & Ratio [95\% CI] \\
\midrule
Short Boolean & 3/3 & 10.74 & 10.66 & 1.01 [0.97, 1.09] \\
Short choice & 3/3 & 10.47 & 10.97 & 0.95 [0.95, 0.96] \\
Three questions & 3/3 & 21.78 & 20.84 & 1.05 [1.03, 1.07] \\
\bottomrule
\end{tabular}

\caption{Liquid versus the released \texttt{reduce-overhead} compiler setting, three completed/planned rounds. Ratio is compiled-native latency divided by DFIP latency; below one favors compilation. Intervals resample paired container medians.}
\label{tab:compiler}
\end{table*}

The earlier eager-baseline ablation isolates the mechanism (Table~\ref{tab:ablation}). Replay alone gives 1.63$\times$; projection alone gives approximately 1.02--1.03$\times$. Replay with either projection gives approximately 1.75$\times$. The small cuBLAS/Triton difference does not support a custom matrix-multiply explanation for the total gain. Encoding, backbone work, output reconstruction, and readout remain timed. The compiler comparison shows why an eager-only headline would overstate incremental benefit.

\begin{table}[t]
\centering\small
\begin{tabular}{lrr}
\toprule
Liquid mode & Median (ms) & Ratio [95\% CI] \\
\midrule
\texttt{native} & 18.69 & 1.00 [1.00, 1.00] \\
\texttt{graphs\_only} & 11.47 & 1.63 [1.18, 1.92] \\
\texttt{candidate\_cublas} & 18.40 & 1.02 [0.95, 1.07] \\
\texttt{candidate\_triton} & 18.12 & 1.03 [1.00, 1.06] \\
\texttt{production\_cublas} & 10.69 & 1.75 [1.26, 2.08] \\
\texttt{production} & 10.68 & 1.75 [1.26, 2.07] \\
\bottomrule
\end{tabular}

\caption{Liquid short Boolean ablation in the earlier eager image, three retained paired rounds. Latencies are geometric means of worker medians. Combined modes use replay and projection; the union contains 40--42 answer tokens. This is separate from Table~\ref{tab:compiler}.}
\label{tab:ablation}
\end{table}

Each round's 128 changing categorical requests preserve choices and sampled threshold classifications for compiler and DFIP relative to eager native. Compilation reaches maximum probability drift 0.0220; DFIP reaches $7.67\times10^{-5}$. These unlabeled diagnostic requests measure agreement with eager native outputs. The audit does not evaluate ground-truth task accuracy, calibration quality, or downstream decision utility; the effect of the observed drift on those outcomes remains unmeasured. Distinct-size capture warnings and admission costs are retained. Neither path is established as bitwise equivalent on every schema.

\subsection{GLiNER: incremental gains over native batching}
The native GLiNER pilot preserves all 29 repeated outputs under the declared comparison. Shared scheduling admits all eighteen schema/size combinations: six ordered schemas at sizes two/four/eight. Primary resident audits preserve labels and confidence. Relative to serial native, shared execution gives 2.92$\times$ for recurring schemas and 1.92$\times$ for diverse schemas. Against stronger native FIFO batching, gains are 1.08$\times$ and 1.57$\times$ (Table~\ref{tab:gliner}). The diverse result is incremental batching evidence in one worker, not a population estimate.

\begin{table*}[t]
\centering\small
\begin{tabular}{llrrrr}
\toprule
Substrate & Trace & Gates & Serial/shared & Batch/shared & Audit \\
\midrule
Eager & homogeneous & 18/18 & 2.92 & 1.08 & pass \\
Eager & diverse & 18/18 & 1.92 & 1.57 & pass \\
Compiled & homogeneous & 18/18 & 2.04 & 1.07 & pass \\
Compiled & diverse & 18/18 & 1.46 & 1.40 & pass \\
\bottomrule
\end{tabular}

\caption{GLiNER resident traces, 96 records each. Each substrate has one worker and five paired randomized warm timings. Ratios divide baseline median by shared-scheduler median. Audits compare exposed labels and confidence. Compilation is a post hoc follow-up with reused fixtures, not independent confirmatory replication.}
\label{tab:gliner}
\end{table*}

The post hoc released-compiler worker also admits all eighteen gates and passes the resident audit. The incremental native-batch ratios are 1.07$\times$ for recurring schemas and 1.40$\times$ for diversity. Admission takes 260.44\,s including first-use compilation, versus 17.98\,s in the eager pilot; those costs are outside the warm medians. The compiler follow-up reuses fixtures and is not an independent confirmation.

GLiNER profiling records host overhead and FP32 matrix operations separately from latency samples; an operator trace is not a full-request estimate. Scheduling exploits compatibility and length variation rather than replacing classifier math. In the compiler follow-up, every mode uses the released compiler substrate and eager outputs from that worker provide audit references. First-use compilation contributes to recorded calibration cost, outside warm timing. The retained 1.40$\times$ diverse native-batch gain warrants independent replication, while its adverse arrival-tail result constrains deployment claims.

\section{Answer Fidelity}
The literal audits each cover 3,024 decisions (Table~\ref{tab:literal}). Decider has zero exposed probability drift, zero changes in 1,112 returned choices, and zero drift in 1,056 scores and 2,168 confidence fields. Clef preserves choices but reaches 0.0188 maximum literal score drift and 0.0129 confidence drift. Probability slots cross 0.1 and 0.25. These are not demonstrated Boolean answer flips, but preclude full probability equivalence.

\begin{table}[t]
\centering\small\setlength{\tabcolsep}{3pt}
\begin{tabular}{lrrr}
\toprule
Checkpoint & Choice changes & Max $|\Delta p|$ & Max $|\Delta s|$ \\
\midrule
PPLX-Decider & 0/1112 & 0.0000 & 0.0000 \\
Clef-Flash & 0/1112 & 0.0129 & 0.0188 \\
\bottomrule
\end{tabular}

\caption{Literal answer audit, one separate native-then-DFIP container per model. Probability drift covers 3,024 decisions, choice changes 1,112 categorical fields, and score drift 1,056 ordinal scores.}
\label{tab:literal}
\end{table}

Serving reveals a different observable effect. Clef retains choices on 2,473 matched successful pairs while reaching 0.0056 maximum probability/confidence drift. Twenty-one pairs cross literal confidence 0.9, with repeated inputs across rounds. These crossings would change a routing policy using that threshold; their effect on task accuracy or downstream routing utility is unmeasured. Unmatched and failed requests remain outside paired audits.

Primary GLiNER resident and serving audits pass with FP32 confidence differences on the order of $10^{-7}$; its compiler audits also pass the declared tolerance. Only exposed fields are tested, not a hidden per-label distribution. Calibration thresholds do not protect an arbitrary threshold closer to a result than the observed drift. Accelerated interfaces should declare preserved fields and test the confidence policy actually consumed by the application.

\section{Changing Arrivals and Serving Limits}
\subsection{Fixed-shape gains reverse under diverse traffic}
Table~\ref{tab:serving} retains observed external phases, including phases from a worker that later fails. Missing phases are not zero-loss successes. Successful-request latency is reported with loss and within-window throughput, so a path cannot appear faster by dropping work.

All three Decider attempts are retained: two complete, and one encounters illegal-memory-access errors in its first graph phase after completing native phases. At five offered requests/s, native loses none and has geometric-mean p95 0.81\,s. Replay gives 2.04 within-window completions/s, 23.0\% loss, and successful-request p95 49.41\,s (range 21.79--76.55\,s). At thirty/s the two observed replay phases lose 83.1\%. All attempts provide 1,761 matched successful pairs with unchanged choices; the two complete graph rounds preserve matched probabilities. This subset cannot certify missing or failed requests.

Recapture explains part of the reversal. One complete attempt records 820 captures and 915 replays including admission and audits, so many replays occur immediately after capture rather than amortizing its cost. At thirty/s its graph service median is 524\,ms versus native 129\,ms. Admission overhead is a bottleneck on this trace; the observations do not locate the fault or establish performance for one recurring shape.

All three Clef serving rounds complete without request errors. At five/s native loses none with p95 0.34\,s; replay loses 4.9\%, completes 2.68/s within the window, and has p95 39.44\,s. At thirty/s replay loses 83.1\% versus serialized native's 54.3\%. This wrapper is not a tuned dynamic-batching ceiling, but the result prevents extrapolating the resident 1.45$\times$ gain to diverse traffic.

\begin{table*}[t]
\centering\small
\begin{tabular}{llrrrrr}
\toprule
Checkpoint & Mode & Offered/s & $n$ & Completed/s & p95 (ms) & Loss (\%) \\
\midrule
PPLX-Decider & Native & 5 & 3 & 4.97 & 810 & 0.0 \\
PPLX-Decider & Native & 15 & 3 & 10.02 & 6998 & 23.4 \\
PPLX-Decider & Native & 30 & 3 & 10.40 & 13393 & 58.3 \\
PPLX-Decider & DFIP replay & 5 & 3 & 2.04 & 49415 & 23.0 \\
PPLX-Decider & DFIP replay & 15 & 2 & 3.36 & 47046 & 63.4 \\
PPLX-Decider & DFIP replay & 30 & 2 & 2.94 & 46422 & 83.1 \\
Clef-Flash & Native & 5 & 3 & 4.98 & 335 & 0.0 \\
Clef-Flash & Native & 15 & 3 & 12.20 & 10078 & 4.5 \\
Clef-Flash & Native & 30 & 3 & 11.59 & 12398 & 54.3 \\
Clef-Flash & DFIP replay & 5 & 3 & 2.68 & 39440 & 4.9 \\
Clef-Flash & DFIP replay & 15 & 3 & 3.48 & 41558 & 62.6 \\
Clef-Flash & DFIP replay & 30 & 3 & 2.94 & 44330 & 83.1 \\
\bottomrule
\end{tabular}

\caption{External serving with 60-second phases. $n$ counts observed container phases, including failed rounds. Completed/s counts within-window responses; p95 is the geometric mean of successful-request container quantiles. Loss includes all drops and errors. Later completions, p99, drain time, and unmatched requests remain archived.}
\label{tab:serving}
\end{table*}

\subsection{Bounded capture does not establish uniform speed}
Separate bounded-admission workers record four successful captures and four resident entries each without request errors. Decider records 474 replays and corrected Clef 400, including audits and immediate post-capture replay. Most calls remain native. These counts verify the budget in those runs, not general CUDA reliability or a same-worker comparison against eviction.

At five/s Decider loses none in either mode. Adaptive p95 is higher on diverse traffic (0.48 versus 0.32\,s), lower on mixed traffic (0.35 versus 0.48\,s). At thirty/s diverse adaptive/native throughput is 9.62/7.53 and loss 60.8/67.8\%; mixed throughput is 8.22/7.65 and loss 65.5/67.4\%. Improvement is modest and uneven with substantial overload. Its 2,057 matched categorical pairs preserve choices, probabilities, and confidence; other schemas and omitted requests remain outside this audit.

Corrected Clef has slightly higher adaptive p95 at five/s on both traces. Both modes sustain 14.93/s without loss at fifteen offered/s. At thirty/s diverse adaptive/native throughput is 15.88/15.95 and mixed 16.55/16.25, with roughly 38--40\% drops. No consistent capacity gain emerges. Its 3,844 matched categorical pairs preserve choices, but probability/confidence drift reaches 0.0065 with fourteen mixed-trace confidence-0.9 crossings. Bounding capture cost does not establish confidence equivalence or general acceleration.

\begin{table*}[t]
\centering\small
\begin{tabular}{lllrrr}
\toprule
Checkpoint & Trace & Mode & Completed/s & p95 (ms) & Loss (\%) \\
\midrule
PPLX-Decider & diverse & adaptive & 4.98 & 484 & 0.0 \\
PPLX-Decider & diverse & native & 4.98 & 318 & 0.0 \\
PPLX-Decider & mixed & adaptive & 4.98 & 348 & 0.0 \\
PPLX-Decider & mixed & native & 4.98 & 476 & 0.0 \\
Clef-Flash & diverse & adaptive & 4.98 & 419 & 0.0 \\
Clef-Flash & diverse & native & 4.98 & 358 & 0.0 \\
Clef-Flash & mixed & adaptive & 4.98 & 349 & 0.0 \\
Clef-Flash & mixed & native & 4.98 & 336 & 0.0 \\
\bottomrule
\end{tabular}

\caption{Separate bounded-admission pilot at five offered requests/s, one validated worker per checkpoint. Diverse precedes mixed traffic with state retained. Decider runs native-first, corrected Clef adaptive-first. No confidence interval or pooling with the earlier serving study is implied.}
\label{tab:adaptive}
\end{table*}

\subsection{GLiNER: resident gains versus arrival tails}
Every primary GLiNER mode completes all 256 requests per trace without drops or errors. The recurring trace produces no multi-request batches at 25/s; shared p95 is 25.56\,ms versus native batching 24.90\,ms and serial 23.15\,ms. Accumulation adds waiting without a batch benefit at this load.

For primary diverse traffic, shared and native batching both have approximately 127.90\,ms p95, versus serial 226.62\,ms. Shared processes 35.9\% of offered requests in multi-request batches versus native batching 32.8\%. The offline median gain therefore does not establish an incremental tail benefit.

The released-compiler follow-up completes all requests without drops or errors, but diverse shared p95 is 72.10\,ms versus compiled native batching 43.17\,ms and compiled serial 69.52\,ms. Thus the incremental resident 1.40$\times$ gain coexists with worse latency under the tested arrivals. Table~\ref{tab:gliner-serving} retains both substrates. These warmed, second-scale resident windows characterize the scheduler, not sustained external-serving reliability.

\begin{table*}[t]
\centering\small
\begin{tabular}{lllrrrr}
\toprule
Substrate & Trace & Mode & p95 (ms) & Complete/offered & Drops & Audit \\
\midrule
Eager & homogeneous & Serial & 23.1 & 256/256 & 0 & pass \\
Eager & homogeneous & Native batch & 24.9 & 256/256 & 0 & pass \\
Eager & homogeneous & Shared & 25.6 & 256/256 & 0 & pass \\
Eager & diverse & Serial & 226.6 & 256/256 & 0 & pass \\
Eager & diverse & Native batch & 127.9 & 256/256 & 0 & pass \\
Eager & diverse & Shared & 127.9 & 256/256 & 0 & pass \\
Compiled & homogeneous & Serial & 19.1 & 256/256 & 0 & pass \\
Compiled & homogeneous & Native batch & 21.3 & 256/256 & 0 & pass \\
Compiled & homogeneous & Shared & 21.8 & 256/256 & 0 & pass \\
Compiled & diverse & Serial & 69.5 & 256/256 & 0 & pass \\
Compiled & diverse & Native batch & 43.2 & 256/256 & 0 & pass \\
Compiled & diverse & Shared & 72.1 & 256/256 & 0 & pass \\
\bottomrule
\end{tabular}

\caption{GLiNER scheduled-arrival replay, 256 offered requests per mode/trace. p95 includes queueing and length estimation and uses completed requests. Drops/errors accompany latency. Compiled rows are a post hoc worker; transport and autoscaling are outside the experiment.}
\label{tab:gliner-serving}
\end{table*}

\section{Related Work}
SGLang~\citep{sglang} supports structured language-model programs and constrained selection; DFIP does not introduce a decision primitive. PagedAttention/vLLM~\citep{vllm} improves generative serving through memory management and batching. Sarathi-Serve~\citep{sarathi} studies throughput--latency tradeoffs using bounded token work and chunked prefills. Our GLiNER policy uses existing batches and a length budget; it neither implements chunked prefills nor replaces generative scheduling. This motivates native-batch comparisons and queueing measurements rather than serial/batch ratios alone.

PyGraph~\citep{pygraph} studies robust CUDA graph compilation, compatibility transformations, and replay overhead. PyTorch documents static-address and control-flow restrictions~\citep{cudagraphs}. Our shape cache and empirical admission are simpler integrations with exposed limitations. Triton~\citep{triton} provides tiled programming; Liquid's ablation shows that reducing projection scope need not appreciably accelerate a full request.

GLiNER2~\citep{gliner2} supplies a schema-driven encoder architecture with classification and task composition. It contextualizes the newer pinned Decide checkpoint; published CPU results are not its implementation identity or our GPU baseline. MLPerf Inference~\citep{mlperf} emphasizes scenarios, quality constraints, and distinct latency/throughput metrics. We follow that separation without claiming compliance. BANKING77~\citep{banking} supplies an independently sourced intent corpus; our typed schemas and ordinal diagnostics are derived workloads.

\section{Limitations and Deployment Implications}
\textbf{Baselines and selection.} Native Qwen kernels and released compilers address specific weaknesses of eager-only comparisons but do not exhaust optimized backends. Four checkpoints selected after development cannot establish unseen-architecture transfer. Prompt and workload differences prevent pooling a cross-model speedup. Clef's missing native round and earlier errors make retained estimates conditional.

\textbf{Serving scope.} Synthetic states emphasize particular paths. External tests use one geography, fixed rates, serialized GPU wrappers, and minute-scale phases. They do not establish a tuned dynamic-batching ceiling, hours-long reliability, or autoscaling. Bounded-admission pilots have one validated worker per model, history-dependent traces, and fixed differing mode orders; no direct eviction-versus-pinned trial isolates the policy effect. GLiNER has one worker per substrate, warmed second-scale replays, and reused fixtures in the post hoc compiler follow-up.

\textbf{Numerical scope.} Rounding, calibration, and reductions can separate label agreement from literal field equality. Five thresholds do not cover arbitrary policies; calibration and subsequent agreement are finite observations. Dataset views correlate, the typed set was used in development, and training overlap is unknown. Native agreement evaluates implementation fidelity. Establishing preservation of task accuracy, calibration quality, and downstream utility beyond the tested samples requires representative, task-specific labeled evaluation.

\textbf{Reliability and cost.} CUDA errors may be reported after the faulting operation. Archived traces cannot assign a kernel cause; synthetic lifetime checks cannot certify full engines. Native fallback cannot repair a poisoned context. GPU-time savings can reduce one compute-cost component, but no billed savings are measured: load, admission, host resources, idle lifetime, queueing, and retries also matter. The evidence supports opt-in workload-specific experiments with native execution as the reviewed default and application confidence policies checked explicitly.

\textbf{Reproducibility.} The public artifact freezes analysis source, a hash-locked CPU analysis environment, and observed GPU dependency constraints. It maps selected results to exact execution commits and retains samples and failed attempts. Historical GPU environments were not fully locked; cloud placement, clocks, contention, unrecorded transitive dependencies, and drivers remain partly uncontrolled.

\section{Conclusion}
DFIP demonstrates workload-specific acceleration for four released System One interfaces using established execution techniques. Selective replay improves repeated short Decider and Clef requests over stronger native kernels, but not the tested diverse serving traffic. Liquid's released compiler closes most of its eager gap; candidate projection alone contributes little. Schema-compatible length-aware batching retains a 1.40$\times$ diverse resident gain over compiled native FIFO batching for GLiNER, while worsening the tested compiled arrival tail. Probability and confidence changes, attrition, and GPU errors define concrete limits. The results favor selecting an execution path per native contract and workload, comparing with efficient native execution, and validating the fields consumed by users. General acceleration and production readiness require independent serving and numerical validation.

\bibliographystyle{plainnat}
\bibliography{references}

\appendix
\section{Artifact and Reproduction}
\label{app:artifact}
The frozen analysis artifact is repository commit \href{https://github.com/superagent-ai/dfip/tree/f2eef1746e5e39b92827759344d4ef47410fc464}{\texttt{\seqsplit{f2eef1746e5e39b92827759344d4ef47410fc464}}}. This is an analysis and evidence revision, not a single execution commit for all experiments or the later manuscript build. From a full clone, check out that revision and reproduce the CPU inputs with:
\begin{quote}\small
\begin{verbatim}
git checkout \
  f2eef1746e5e39b92827759344d4ef47410fc464
python3.12 -m venv .venv-analysis
. .venv-analysis/bin/activate
python -m pip install \
  --require-hashes \
  -r repro/analysis-py312.lock
python scripts/verify_artifacts.py \
  --git-history --regenerate
\end{verbatim}
\end{quote}
The verification command checks archived evidence and reproduces the focused summary and ten TeX inputs, including Table~\ref{tab:overview}; it does not rerun GPU timings. The artifact's \texttt{repro/README.md} and the current \texttt{docs/reproducibility.md} explain the separate GPU reconstruction steps.

The per-run index \texttt{repro/source-index.json} maps all 30 selected raw records to their recorded and resolved full Git commits, SHA-256 hashes, and observed environments. For example, Decider's strengthened resident rounds use \texttt{\seqsplit{731e3f26aca477d4726010345dae724bc13d3c4b}} and Clef's use \texttt{\seqsplit{b7768962a93c43ef18230163d0079ece17708357}}. The readout, external-serving, and Liquid compiler phase uses \texttt{\seqsplit{bef713cac4a57a3f1092abc5d7cb35d63a3399c5}}; GLiNER's compiler follow-up uses \texttt{\seqsplit{4adadc36c6cc2e0cd95dad0b6c67a07462649886}}. Other phase commits, including corrected or excluded attempts, remain in the index. Reconstruct GPU trials from their own source snapshots and protocols, rather than substituting the later analysis revision.

Observed fast-image constraints pin Torch 2.14.0, Transformers 5.17.0, Triton 3.8.0, NumPy 2.5.4, tokenizers 0.23.3, causal-conv1d 1.7.0, fla-core 0.5.2, and other recorded packages. The separate GLiNER constraints retain its six recorded package versions, including gliner2 2.1.0 and peft 0.21.2. These profiles are evidence-backed pip constraints, not complete historical container locks. Unrecorded GLiNER transitive versions are not inferred from another image. The CPU analysis lock is newly validated and distinct from the original GPU environments. A rebuilt GPU environment is a new experimental condition; record its resolved dependencies and results separately.

The repository contains adapters, remote preparation, pinned checkpoint registries, frozen specifications, source snapshots, timing and answer records, failure excerpts, and deterministic analysis. Focused tables are generated from retained evidence, not estimated values. Original trials for the four checkpoints remain available with exclusions annotated.

Reproduction requires a Modal workspace and checkpoint access. The client keeps weights and corpora remote. Run one study application at a time with at most one GPU worker. Optimized-native, compiler, literal-audit, and external-serving results are separate from earlier component ablations. \texttt{scripts/analyze\_strengthened.py} verifies accounting and worker identity. \texttt{scripts/analyze\_adaptive.py} verifies budgets and retains Clef's excluded replacement attempt. \texttt{scripts/analyze\_contracts.py} recomputes GLiNER resident comparisons and checks arrival/batch accounting. \texttt{scripts/analyze\_focused.py} regenerates this paper's four-model tables, plot, and evidence manifest without pooling protocols.

GLiNER's original pilot retains calibration summaries and resident references and outputs. Full individual calibration responses and complete serving references were added for the compiler follow-up. Its specification declares post hoc reuse and one worker. Commit-named snapshots and journals identify code per trial. No script is a production certification procedure.

\section{Checkpoint and Dataset Identity}
The pinned revisions are Liquid d1-3B \texttt{\seqsplit{051bcc464b01b9f92942b364d9586b0ef5912432}}, Clef-Flash \texttt{\seqsplit{8b2e5fd17c09fd49bc2880b805d5323ec7ff4ff3}}, PPLX-Decider \texttt{\seqsplit{6195aa55a72ae21e503c8a60198f7b390d42adb5}}, and GLiNER2.5-Decide \texttt{\seqsplit{9d1bfb848cd16d93ee56ccbc59cd92bf0f284d90}}. Typed corpus revision is \texttt{\seqsplit{c76749ec58bd8c3d2ea706b31c333a9059c38f90}}; BANKING77 is \texttt{\seqsplit{8eed2b348ec07d40ffbe91dcc4718e2f0f977714}}. Dataset loaders verify recorded corpus checksums; derived BANKING77 records use seed 20261012. Checkpoint revisions are pinned, archived source and result files carry SHA-256 hashes, and released verification is retained where available. Each trial's manifest is authoritative rather than assuming one shared image.

The source builds with \texttt{latexmk -pdf main.tex}. A CPU-only Modal entry point is provided without local TeX. The accompanying artifact includes source, generated tables/figure, and evidence.
\end{document}